\documentclass[pra,twocolumn,superscriptaddress]{revtex4-2}
\usepackage{amssymb}
\usepackage{amsfonts}
\usepackage{amsmath}
\usepackage{bbm}
\usepackage{mathrsfs}
\usepackage{graphics,graphicx,epsfig,bm,amsmath,amsthm,amssymb}
\usepackage{bm}
\usepackage{bbm}
\usepackage{longtable}
\usepackage{multirow}
\usepackage{array}
\usepackage{color}
\usepackage[usenames,dvipsnames]{xcolor}
\usepackage[right]{lineno}
\newcommand{\ket}[1]{\left| #1\right\rangle}  
\usepackage{float}
\usepackage[colorlinks=true,
linkcolor=blue,citecolor=blue,
pdfauthor={ },
pdftitle={ },
pdfsubject={ },
pdfkeywords={ }]{hyperref}

\begin{document}
\title{Electric imaging and dynamics of photo-charged graphene edge}
	
\author{Zhe Ding}
\email{These authors contributed equally to this work.}
\affiliation{CAS Key Laboratory of Microscale Magnetic Resonance and School of Physical Sciences, University of Science and Technology of China, Hefei, 230026, China.}
\affiliation{Anhui Province Key Laboratory of Scientific Instrument Development and Application, University of Science and Technology of China, Hefei, 230026, China.}
\affiliation{Hefei National Laboratory, University of Science and Technology of China, Hefei, 230088, China.}

\author{Zhousheng Chen}
\email{These authors contributed equally to this work.}
\affiliation{CAS Key Laboratory of Microscale Magnetic Resonance and School of Physical Sciences, University of Science and Technology of China, Hefei, 230026, China.}
\affiliation{Anhui Province Key Laboratory of Scientific Instrument Development and Application, University of Science and Technology of China, Hefei, 230026, China.}
\affiliation{Department of Chemical Physics, University of Science and Technology
of China, Hefei, 230026, China.}

\author{Xiaodong Fan}
\affiliation{CAS Key Laboratory of Strongly-Coupled Quantum Matter Physics,
and Department of Physics , University of Science and Technology of
China, Hefei, 230026, China.}

\author{Weihui Zhang}
\affiliation{CAS Key Laboratory of Strongly-Coupled Quantum Matter Physics,
and Department of Physics , University of Science and Technology of
China, Hefei, 230026, China.}

\author{Jun Fu}
\affiliation{Hefei National Laboratory, University of Science and Technology of China, Hefei, 230088, China.}
\affiliation{International Center for Quantum Design of Functional Materials
(ICQD), Hefei National Research Center for Physical Science at the
Microscale , University of Science and Technology of China, Hefei,
230026, China.}

\author{Yumeng Sun}
\affiliation{CAS Key Laboratory of Microscale Magnetic Resonance and School of Physical Sciences, University of Science and Technology of China, Hefei, 230026, China.}
\affiliation{Anhui Province Key Laboratory of Scientific Instrument Development and Application, University of Science and Technology of China, Hefei, 230026, China.}

\author{Zhi Cheng}
\affiliation{CAS Key Laboratory of Microscale Magnetic Resonance and School of Physical Sciences, University of Science and Technology of China, Hefei, 230026, China.}
\affiliation{Anhui Province Key Laboratory of Scientific Instrument Development and Application, University of Science and Technology of China, Hefei, 230026, China.}

\author{Zhiwei Yu}
\affiliation{CAS Key Laboratory of Microscale Magnetic Resonance and School of Physical Sciences, University of Science and Technology of China, Hefei, 230026, China.}
\affiliation{Anhui Province Key Laboratory of Scientific Instrument Development and Application, University of Science and Technology of China, Hefei, 230026, China.}

\author{Kai Yang}
\affiliation{CAS Key Laboratory of Microscale Magnetic Resonance and School of Physical Sciences, University of Science and Technology of China, Hefei, 230026, China.}
\affiliation{Anhui Province Key Laboratory of Scientific Instrument Development and Application, University of Science and Technology of China, Hefei, 230026, China.}
\affiliation{Hefei National Laboratory, University of Science and Technology of China, Hefei, 230088, China.}

\author{Yuxin Li}
\affiliation{CAS Key Laboratory of Microscale Magnetic Resonance and School of Physical Sciences, University of Science and Technology of China, Hefei, 230026, China.}
\affiliation{Anhui Province Key Laboratory of Scientific Instrument Development and Application, University of Science and Technology of China, Hefei, 230026, China.}

\author{Xing Liu}
\affiliation{Shanghai Applied Radiation Institute and State Key Lab, Advanced
Special Steel, Shanghai University, Shanghai, 200444, China.}

\author{Pengfei Wang}
\affiliation{CAS Key Laboratory of Microscale Magnetic Resonance and School of Physical Sciences, University of Science and Technology of China, Hefei, 230026, China.}
\affiliation{Anhui Province Key Laboratory of Scientific Instrument Development and Application, University of Science and Technology of China, Hefei, 230026, China.}
\affiliation{Hefei National Laboratory, University of Science and Technology of China, Hefei, 230088, China.}

\author{Ya Wang}
\affiliation{CAS Key Laboratory of Microscale Magnetic Resonance and School of Physical Sciences, University of Science and Technology of China, Hefei, 230026, China.}
\affiliation{Anhui Province Key Laboratory of Scientific Instrument Development and Application, University of Science and Technology of China, Hefei, 230026, China.}
\affiliation{Hefei National Laboratory, University of Science and Technology of China, Hefei, 230088, China.}

\author{Jianhua Jiang}
\affiliation{CAS Key Laboratory of Microscale Magnetic Resonance and School of Physical Sciences, University of Science and Technology of China, Hefei, 230026, China.}

\author{Hualing Zeng}
\affiliation{Hefei National Laboratory, University of Science and Technology of China, Hefei, 230088, China.}
\affiliation{International Center for Quantum Design of Functional Materials
(ICQD), Hefei National Research Center for Physical Science at the
Microscale , University of Science and Technology of China, Hefei,
230026, China.}

\author{Changgan Zeng}
\affiliation{Hefei National Laboratory, University of Science and Technology of China, Hefei, 230088, China.}
\affiliation{CAS Key Laboratory of Strongly-Coupled Quantum Matter Physics,
and Department of Physics , University of Science and Technology of
China, Hefei, 230026, China.}

\author{Guosheng Shi}
\affiliation{Shanghai Applied Radiation Institute and State Key Lab, Advanced
Special Steel, Shanghai University, Shanghai, 200444, China.}

\author{Fazhan Shi}
\email{fzshi@ustc.edu.cn}
\affiliation{CAS Key Laboratory of Microscale Magnetic Resonance and School of Physical Sciences, University of Science and Technology of China, Hefei, 230026, China.}
\affiliation{Anhui Province Key Laboratory of Scientific Instrument Development and Application, University of Science and Technology of China, Hefei, 230026, China.}
\affiliation{Hefei National Laboratory, University of Science and Technology of China, Hefei, 230088, China.}
\affiliation{School of Biomedical Engineering and Suzhou Institute for Advanced Research, University of Science and Technology of China, Suzhou, 215123, China.}

\author{Jiangfeng Du}
\affiliation{CAS Key Laboratory of Microscale Magnetic Resonance and School of Physical Sciences, University of Science and Technology of China, Hefei, 230026, China.}
\affiliation{Anhui Province Key Laboratory of Scientific Instrument Development and Application, University of Science and Technology of China, Hefei, 230026, China.}
\affiliation{Hefei National Laboratory, University of Science and Technology of China, Hefei, 230088, China.}
\affiliation{Institute of Quantum Sensing and School of Physics, Zhejiang
University, Hangzhou, 310027, China.}

 
\begin{abstract}
The one-dimensional side gate based on graphene edges shows a significant capability of reducing the channel length of field-effect transistors, further  increasing the integration density of semiconductor devices. The nano-scale electric field distribution near the edge provides the physical limit of the effective channel length, however, its imaging under ambient conditions still lacks, which is a critical aspect for the practical deployment of semiconductor devices. Here, we used scanning nitrogen-vacancy microscopy to investigate the electric field distribution near edges of a single-layer-graphene. Real-space scanning maps of photo-charged floating graphene flakes were acquired with a spatial resolution of $\sim$ 10 nm, and the electric edge effect was quantitatively studied by analyzing the NV spin energy level shifts due to the electric Stark effect. Since the graphene flakes are isolated from external electric sources, we brought out a theory based on photo-thermionic effect to explain the charge transfer from graphene to oxygen-terminated diamond probe with a disordered distribution of charge traps. Real-time tracing of electric fields detected the photo-thermionic emission process and the recombination process of the emitted electrons. This study provides a new perspective for graphene-based one-dimensional gates and opto-electronics with nanoscale real-space imaging, and moreover, offers a novel method to tune the chemical environment of diamond surfaces based on optical charge transfer.    
\end{abstract}

	
	
	
	\maketitle
	

When exploring the next-generation semiconductor devices, ``one-dimensional gate" technology shows a significant capability of reducing the gate length of field-effect transistors (FET), becoming a key factor in further advancing Moore's Law. This technology leverages the unique properties of low-dimensional materials, such as one-dimensional edges and high electronic control capabilities, providing the possibility for manufacturing smaller and more efficient electronic devices \cite{desaiMoS2Transistors1nanometer2016, wuVerticalMoS2Transistors2022, ahnIntegrated1DEpitaxial2024}. The configuration of using the edge of a single-layer graphene as a side gate can reduce the gate length to below 1 nm, and due to the high electrical conductivity of graphene, such devices have a high On/Off ratio, making them a strong candidate for the next generation of FETs \cite{wuVerticalMoS2Transistors2022}.

Nanoscale real-space electric field imaging near such one-dimensional gates can provide crucial information for the design of FETs. Unfortunately, the widely used Kelvin Probe Force Microscope \cite{kimRewritableGhostFloating2017} can only scan the gate surface potential, while not able to provide direct information about the nearby electric field distribution. Transmission Electron Microscope \cite{ishikawaDirectElectricField2018, coupinMappingNanoscaleElectrostatic2023} and Single Electron Transistor \cite{martinObservationElectronHole2008, ellaSimultaneousVoltageCurrent2019}, on the other hand, are limited in their application to practical semiconductor devices due to the strict experimental environment and the difficulties in sample preparation and probe fabrication. The Nitrogen-Vacancy (NV) color center in diamond is a type of atom-like defect, which has been working as a single-spin scanning probe \cite{balasubramanianNanoscaleImagingMagnetometry2008, maletinskyRobustScanningDiamond2012,duSinglemoleculeScaleMagnetic2024}. Such a probe has been applied to the study of various condensed matter systems, including single-spin systems \cite{grinoldsNanoscaleMagneticImaging2013}, antiferromagnetic materials \cite{grossRealspaceImagingNoncollinear2017, hedrichNanoscaleMechanicsAntiferromagnetic2021, dingObservationUniaxialStrain2023}, two-dimensional ferromagnetic materials \cite{thielProbingMagnetism2D2019, songtianchengDirectVisualizationMagnetic2021}, and Dirac fluids \cite{kuImagingViscousFlow2020, voolImagingPhononmediatedHydrodynamic2021a, jenkinsImagingBreakdownOhmic2022}. Notably, based on the electric Stark effect, NV centers as electric quantum sensors have been studied since 2011 \cite{doldeElectricfieldSensingUsing2011, liNanoscaleElectrometryBased2020}, and recently, scanning electric imaging technology has been developed \cite{bianNanoscaleElectricfieldImaging2021,qiuNanoscaleElectricField2022,palmObservationCurrentWhirlpools2024}, demonstrating the ability to image electric field gradients. Scanning NV Microscopy (SNVM) at room temperature, as a non-invasive method for directly measuring the electric field, is suitable for direct imaging of the electric distribution near one-dimensional gate candidates like graphene edges. Additionally, diamond has a broad transmission spectrum, which allows for local photo-excitation of graphene during the imaging process to study the photo-response process.


This work employed SNVM to study the electric field distribution near a floating graphene's edge under ambient conditions. A nanoscale electric field scanning map with a spatial resolution of $\sim 10$ nm showed evident field enhancing near the graphene edge. A quantitative analysing of scanning profiles across the graphene edge shows a potential difference of $\sim$ 0.81 V between the graphene and silicon substrate. Here, the optical charge transfer process is interesting due to the absence of a connection between the floating graphene and an external electric source \cite{kimRewritableGhostFloating2017}. By locally exciting the sample with a focused 532 nm continuous-wave laser, the graphene showed a charged state revealed by the aforementioned scanning map. According to previous studies, there are disordered distributed charge traps on oxygen-terminated diamond surfaces \cite{sangtawesinOriginsDiamondSurface2019, dwyerProbingSpinDynamics2022}. Electrons can hop among these charge traps, providing a diffusion path for photo-thermionically emitted electrons from graphene. Due to the Coulomb interaction, the diffusion ultimately ceases, and the diamond-graphene system forms a steady state driven by the continuous laser. Such a steady state has a lifetime of several minutes after the continuous laser is turned off and can be traced in real time with an NV probe. 
The phenomena observed in this work can serve as candidates for one-dimensional light-controlled floating gates \cite{massicottePhotothermionicEffectVertical2016, wuVerticalMoS2Transistors2022}, and moreover, this electron transfer process also plays a significant role in catalysis with graphene. Combined with the high sensitivity of NV to magnetic fields, this technology can be used to study electromagnetic noise and microscopic transport phenomena under photo-excitation at the edges of graphene, deepening the understanding of low-dimensional systems. 
	

\begin{figure*}
	\centering
	\includegraphics[width=1\linewidth]{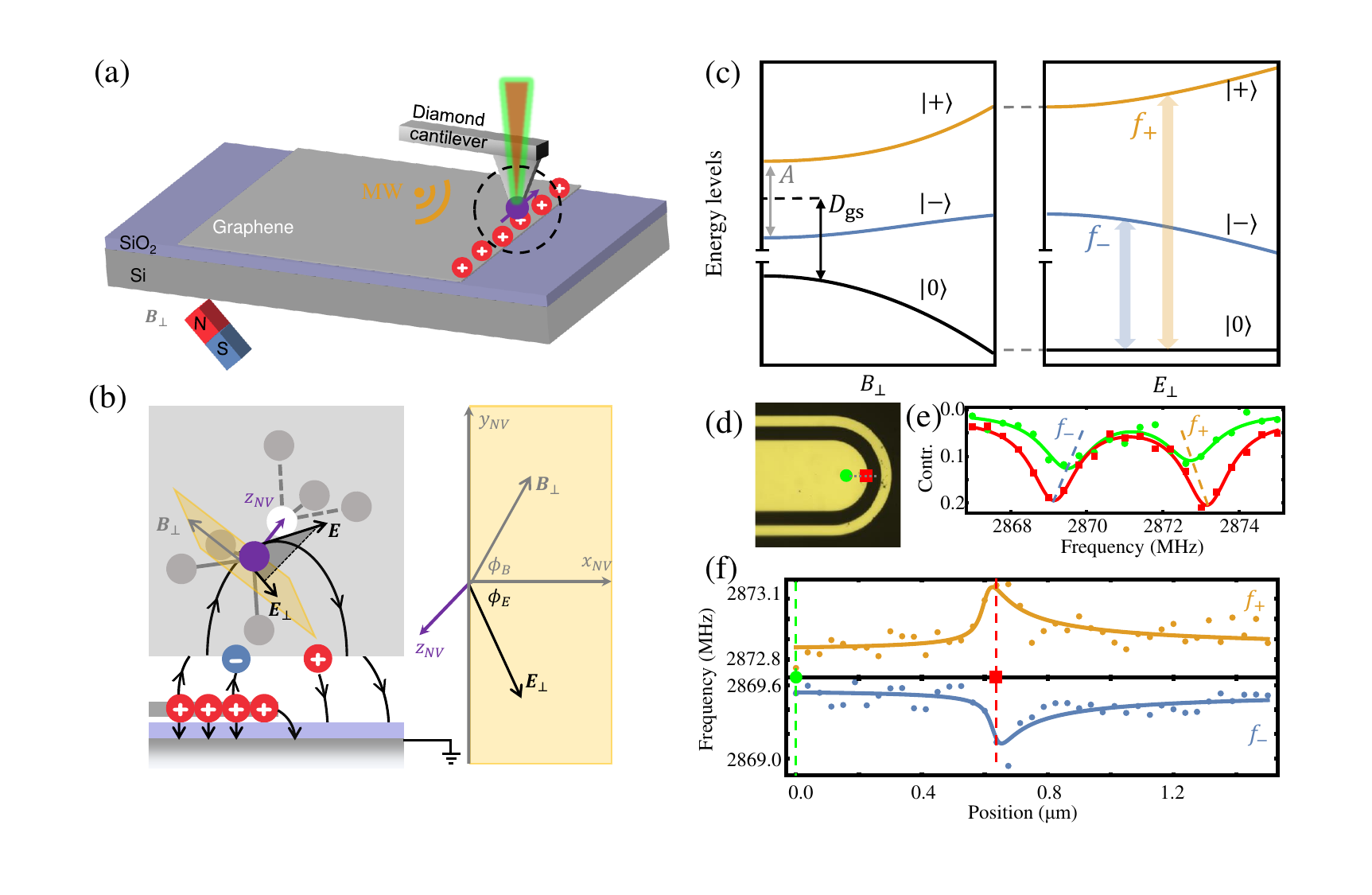}
	\caption{Principle of Electric Scanning NV Microscopy. (a) Configuration of an SNVM imaging electric field at the edge of a floating SLG. The purple spin is the NV center while the red cone with a green aura indicates the optical detection process of the spin state. (b) Details of NV electric detection. The left side is an enlarged view of the black dashed circle in (a). The electric field generated by the charges at the edge of the graphene is represented by black electric field lines. The light gray area represents the diamond probe with a structural diagram of an NV center inside, with the dark gray sphere, purple sphere and white sphere represent carbon, nitrogen atom and lattice vacancy respectively. The purple arrow connecting the nitrogen atom and vacancy indicates the NV axis. The light yellow plane perpendicular to the NV axis is shown in detail in the right part. (c) Spin energy levels of NV under an external perpendicular magnetic field and electric field. (d) Electrode used for the demonstration of scanning NV imaging for electric edging effect. (e) ODMR spectra away from (green round points) and close to (red square points) the electrode edge. (f) Spatial scanning result of NV resonance frequencies across the electrode edge shown in (d). The green circle and red square correspond to markers in (d) and the orange and blue data lines labeled with $f_\pm$ correspond to peaks in (e) and quantum transitions in (c). An edge electric field profile near conductor corners \cite{jacksonClassicalElectrodynamicsThird1998} gives the fitted solid curves.  }
	\label{fig:NVprinciple}
\end{figure*}

A home-built SNVM was used to perform electric imaging near a single-layer-graphene (SLG), as shown in Fig. \ref{fig:NVprinciple} (a). Here, an exfoliated SLG was transferred to a SiO$_2$(285 nm)/Si substrate and the electric field near its edge was imaged by scanning the diamond probe with an NV center at the tip, while its structure is depicted in Fig. \ref{fig:NVprinciple}(b). The NV electronic ground state possesses a spin of $S=1$ while the spin state initialization, operation, and readout can be achieved using Optically Detected Magnetic Resonance (ODMR) technique. In our experiment, ODMR spectra were acquired by collecting NV photon counts while scanning microwave (MW) frequency. Thus, the shift of NV spin energy levels under an electric field can be detected by fitting resonant frequencies, as shown in Fig. \ref{fig:NVprinciple} (c) and (e). 
	

The resonant frequencies detected in ODMR experiments characterize the NV energy level splitting due to the relation $f_\pm=E_{\ket{\pm}} - E_{\ket{0}}$, further enabling electric detection. Here, by decomposing the electric and magnetic field into axial ($B_z, E_z$) and transversal components ($\mathbf{B}_\perp, \mathbf{E}_\perp$) according to the NV axis, as shown in Fig. \ref{fig:NVprinciple} (b), where the azimuth angles of $\mathbf E$ and $\mathbf B$, $\phi_E$ and $\phi_B$, are also labeled in the $x_{\text{NV}}-y_{\text{NV}}$ plane. Former researches have shown that a high axial magnetic field $B_z$ suppresses the electric effect to second order \cite{doldeElectricfieldSensingUsing2011,huxterImagingFerroelectricDomains2023}. So, before the experiments, $B_z$ had been carefully adjusted to zero as shown in Section I and V of the Supplemental Material. In such a configuration, NV energy levels can be written as following: 
\begin{equation}
	\left\{ 
	\begin{aligned}
		&E_{\ket{0}}=-\frac{(\gamma_{e} B_{\perp})^2}{D_{\mathrm{gs}}} \\
		&E_{\ket{\pm}}= D_{\mathrm{gs}}+\frac{(\gamma_{e} B_{\perp})^2}{2D_{\mathrm{gs}}}\pm F(\mathbf{B},\mathbf{E})
	\end{aligned}
	\right. ,
\end{equation}
in which \\
$$
\begin{aligned}
    F(\mathbf{B},\mathbf{E}) =& \left[ \left(\frac{A}{2} \right)^2 +(d_{\perp}E_{\perp})^2 + \frac{(\gamma_{e}B_{\perp})^{4}}{4D_{\text{gs}}^{2}} \right. \\
    & \left. -d_{\perp}E_{\perp} \frac{(\gamma_{e}B_{\perp})^2}{D_{\text{gs}}}\sin(\phi_{E}+2\phi_{B}) \right]^{1/2}.
\end{aligned}
$$
	
In this equation, the zero-field splitting $D_\text{gs}\approx 2870$ MHz may deviate slightly from the theoretical result due to axial electric field or local strain. The electric transversal coupling parameter $d_{\perp}=17$ Hz cm V$^{-1}$  comes from the electric Stark effect \cite{doldeElectricfieldSensingUsing2011}. The gyromagnetic ratio of electron spin $\gamma_e=2.8$ MHz/Gauss and hyperfine parameter $A=3.03$ MHz are related to the magnetic Zeeman effect and hyperfine coupling with the $^{15}$N nuclear spin respectively \cite{dohertyNitrogenvacancyColourCentre2013}. According to the results, NV energy levels under transversal magnetic fields $B_\perp$ and electric fields $E_\perp$ are shown in Fig. \ref{fig:NVprinciple} (c). 
	
We scanned the edge of an electrode to demonstrate the electric imaging capability of SNVM. Fig. \ref{fig:NVprinciple} (d) shows an optical microscopy photograph of the electrode. The scanning path of the NV probe is marked with a gray dotted line, and the real-space scanning result is shown in Fig. \ref{fig:NVprinciple} (f), where the orange and blue curves represent the spatial variation of two ODMR resonance frequencies, $f_+$ and $f_-$, respectively. The spatial positions marked by the green dot and red square correspond to the markers in (d). It is evident that due to the electric edge effect, there is a significant change in the resonance frequency at the edge of the electrode. The ODMR spectra lines far from (green dots) and close to (red squares) the edge are displayed in Fig. \ref{fig:NVprinciple} (e). Notably, the spectral contrast is significantly enhanced when the NV approaches the edge. This may be due to the enhanced electric field, which increases the probability of the NV$^-$ charge state by adjusting the energy bands close to the diamond surface \cite{bianNanoscaleElectricfieldImaging2021}. 
	
\begin{figure*}[tph]
	\centering
	\includegraphics[width=0.8\linewidth]{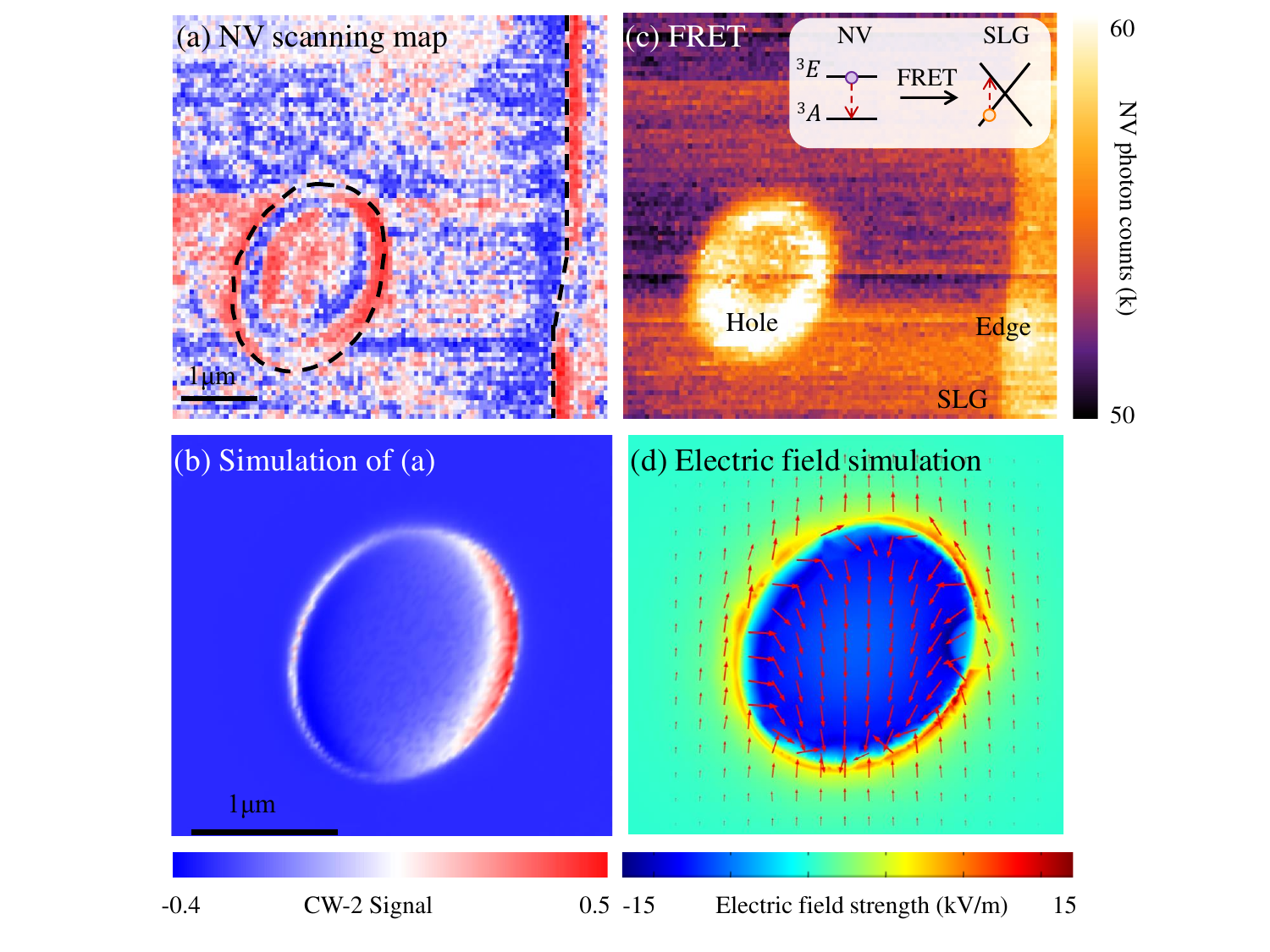}
	\caption{Real-space image of a single-layer-graphene. (a) CW-2 scanning of a SLG near the edge. A hole burnt with a femto-second laser pulse is to the left of the edge. (b) A simulated CW-2 scanning result of an oval hole with the same size of the burnt hole in (a). (c) NV photon counts scanning at the same area. The inset shows the principle of FRET, in which the energy of excited NV center is transferred to an exciton in graphene. (d) Simulation results of the electric field near the hole. The color scale represents the field strength and the arrows represent the in-plane field direction and strength.  }
	\label{fig:ElectricImage}
\end{figure*}

After demonstrating the electric imagining capability, such an SNVM configuration was utilized to study the electric edge effect of charged floating SLG flakes. The samples imaged during the experiments were mechanically exfoliated from highly oriented pyrolytic graphite as shown in Section II of the Supplemental Material. To enhance the imaging speed, we adopted a reduced sampling mode, CW-2,  with the protocol details displayed in Section IV of the Supplemental Material. This section also proves the proportional relation between the CW-2 signal and the resonant frequency shift when it is small compared to the peak width. The electric field scanning results are shown in Fig. \ref{fig:ElectricImage} (a), in which the electric edge effect of a hole and a straight edge are visible.  
	
	Short-range fluorescence resonance energy transfer (FRET) has been demonstrated to exist between the NV excited state and the graphene exciton \cite{tislerSingleDefectCenter2013}. A brief picture of the FRET process is shown in the inset of Fig. \ref{fig:ElectricImage} (c), during which the excited NV center is quenched to ground state with an exciton created in the SLG. According to F\"orster’s law, the quenching rate is given by
	\begin{equation}
		\gamma_{\text{nr}}=\gamma_{\mathrm{r}} \frac{z_{0}^{4}}{z^{4}}, \label{forster}
	\end{equation}
	where $\gamma_{\mathrm{r}}\approx 117.6\text{ MHz}$ is the decay rate of the excited NV when it is away from SLG flakes, and $z_0=15.3\text{ nm}$ has been demonstrated by Tisler et. al \cite{tislerSingleDefectCenter2013}. This formula shows that the topography of graphene flakes can be imaged by extracting NV photon counts from the real-space scanning data, as shown in Fig. \ref{fig:ElectricImage} (c). On the right side of the image is a natural boundary, it is evident that the NV photon count in SLG area is lower due to FRET compared to the external regions. On the left side of the image, there is an oval-shaped bright area, which corresponds to a hole burnt by focused femto-second laser pulses, and its SEM photograph can be found in Section II of the Supplemental Material. Both Fig. \ref{fig:ElectricImage} (c) and the SEM photograph show that there is a circular island inside the oval area, which should be a type of graphene nano-crystallite generated by the femto-second laser pulse \cite{robertsResponseGrapheneFemtosecond2011}. Since the photon count of NV close to the island is higher than that of the SLG, the FRET efficiency of the island is lower, which is consistent with the properties of the nano-crystallite.
	
	From the photon counts map, the outlines of the edge and hole are extracted and drawn in Fig. \ref{fig:ElectricImage} (a) for guidance. An evidently enhanced electric field due to the edge effect arises near such outlines. The CW-2 map of such an oval-shaped hole is simulated and displayed in Fig. \ref{fig:ElectricImage}(b), the simulation indicates that the inhomogeneity of the signal along the hole edge arises from the directional asymmetry introduced by the NV axis and the transverse magnetic field. Fig. \ref{fig:ElectricImage} (d) gives an example of  the simulated electric field distribution when the diamond probe is to the right edge of the hole. 
	
	
	\begin{figure*}[tph]
		\centering
		\includegraphics[width=0.8\linewidth]{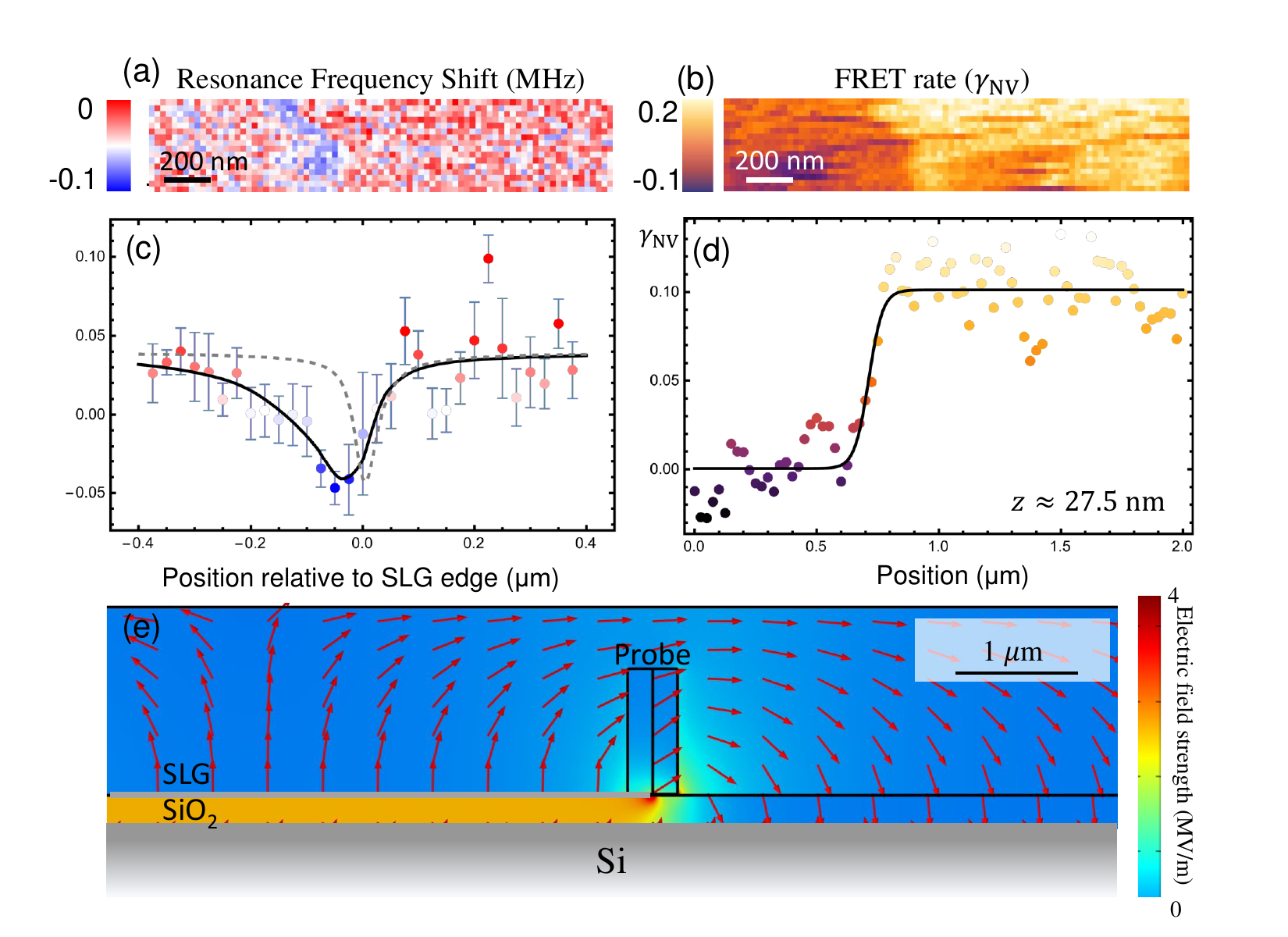}
		\caption{Quantitative analysis of electric field distribution near the graphene edge. (a) Real-space scanning result of spin resonance frequency shift. (b) Real-space scanning result of NV FRET rate. (c) Spin resonant frequencies across the edge. The solid black line represents the fitting result using the charged graphene model, while the gray line is the line shape calculated from the electric dipole layer model with the same NV configuration.(d) Fitting of FRET rate along the first scanning line in (b), as an example of locating edge positions with FRET rate. (e) Simulation results of electric field near the edge.   }
		\label{fig:ElectricAnalysis}
	\end{figure*}
	
	To study the edge effect on the electric field distribution of charged graphene with higher confidence, a fine scanning at the edge was conducted, as shown in Fig. \ref{fig:ElectricAnalysis} (a), in which a full ODMR spectrum was performed to acquire NV resonance frequency for each pixel. Fig. \ref{fig:ElectricAnalysis} (a) displays the shift relative to the resonance frequency under zero electric field (2869.5 MHz). Moreover, the FRET-induced quenching rate of NV excited state, $\gamma_{\text{nr}}$, shown in Fig. \ref{fig:ElectricAnalysis} (b), can be obtained from the distribution $I(x,y,z)$ of the NV photon count rate using the formula
	\begin{equation}
		\gamma_{\mathrm{nr}}=\left(\frac{I_{0}-I_{\mathrm{B}}}{I(x, y, z)-I_{\text{B}}}-1\right) \gamma_{\text{r}}
	\end{equation}
	where $I_{0}$ and $I_{\text{B}}$ are the NV photon count rate far from the SLG and the background photon count rate (including stray photon and the dark count rate of the photon detector), respectively. 
	
	Multiple rows of ODMR data in Fig. \ref{fig:ElectricAnalysis} (a) can be accumulated by translating the edge positions of each row to the origin with the help of aforementioned quenching rate imaging, thereby improving the signal-to-noise ratio. An example of edge determination is shown in Fig. \ref{fig:ElectricAnalysis} (d) \footnote{Analyzing the FRET rate for each scanning line provides crucial information such as the position of the SLG edge and the NV-sample distance.  Taking the first line as an example, Fig. \ref{fig:ElectricAnalysis} (d) plots $\gamma_{\text{nr}}$ in the unit of $\gamma_{\text{r}}$ and the fitting result is shown by the black curve. The fitting result shows that $\gamma_{\text{nr}}(\text{on SLG})\approx 0.10\ \gamma_{\text{r}}$ and the NV-SLG distance $z=27.2 \text{ nm}$ is calculated with the help of equation (\ref{forster}). Additionally, fitting the spatial distribution's rising edge of the FRET rate yields the position of the SLG edges, which provides more information for analysis of the electric edge effect.}
 The quenching rate gives a probe-sample distance of 27.5 nm, which indicates a spatial resolution of $\sim$ 10 nm scale \cite{casolaProbingCondensedMatter2018}. Such an electric profile across the edge is plotted in Fig. \ref{fig:ElectricAnalysis} (c), in which the standard deviation of multiple rows is used to provide error bars. Using the COMSOL finite element analysis method, the electric field distribution of the system is simulated, Fig. \ref{fig:ElectricAnalysis} (e) shows the simulation results when the diamond probe coincides with the edge of the graphene. By running such simulations with the NV probe position scanned, we extracted the goal function that the electric profile in Fig. \ref{fig:ElectricAnalysis} (c) should follow. The experimental results are fitted and shown by the black curve. The fitting results indicate a relative potential difference between the graphene and the Si substrate of $0.81 \text{ V}$, which gives an average electric field strength of 28 kV/cm in SiO$_2$.

	\begin{figure}[tph]
		\centering
		\includegraphics[width=1\linewidth]{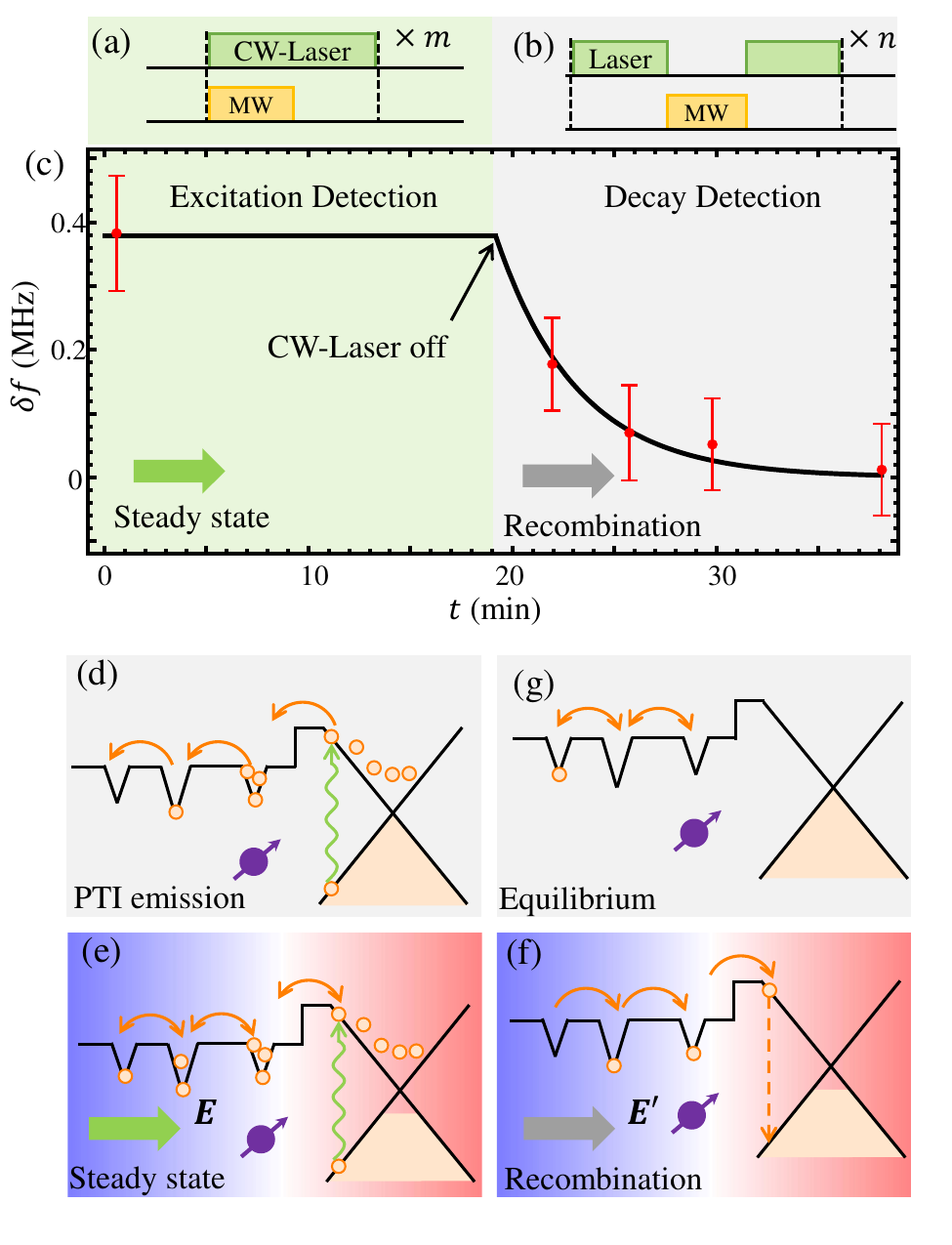}
		\caption{Real-time trace of the photo-thermionic diffusion and recombination. (a) ODMR sequence diagram for the excitation detection. During the detection, a green laser is continuously applied (CW-Laser), and a MW pulse with a length of 1 ms is applied periodically, during which photon counts are detected. (b) ODMR sequence diagram for the decay detection. The first laser pulse with a length of 1 $\mu$s initialize the NV spin into $|0\rangle$, after which a MW pulse is applied to rotate the NV spin. A second laser pulse with a length of 1 $\mu$s is applied at last for spin state detection. The sequences in (a) and (b) are repeated for $m=10^3$ and $n=10^6$ times to acquire enough photon counts. (c) Real-time trace of the photo-doping decay. The NV spin resonant frequency shift $\delta f$'s evolution is plotted and fitted with an exponential decay model. The excitation and decay detection results corresponding to diagrams (a) and (b), respectively, are marked with green and gray background.(d-g) Micro-mechanism of PTI emission, steady state, recombination, and equilibrium, respectively. }
		\label{fig:RealtimeTrace}
	\end{figure} 
	
Noting that the charged SLG flake was a floating gate with no electric connection to any external source, the charge transfer process is of interest \cite{kimRewritableGhostFloating2017}. 	To study the dynamics of such a process, a real-time tracing experiment observed the NV electric signal during and after this process. As shown in Fig. \ref{fig:RealtimeTrace} (a) and (c), a CW experiment measures the electric field at the edge of a charged SLG flake. After the continuous laser was turned off, a sequence of pulse experiments traced the edge electric field signal in real time, as shown in Fig. \ref{fig:RealtimeTrace} (b), the results of pulse experiments are labeled Pulse-2, with its protocol discussed in Section IV of the Supplemental Material. The real-time data show that after the continuous laser is turned off, a discharge process of SLG and ultimately causes the pulse ODMR signal's exponential decaying, which reveals a lifetime of $\sim 4$ min. 
 
To explain the real-time tracing results, the charge and discharge pathway is considered here. Though Coulomb traps in SiO$_2$ have been reported to be capable of accepting transferred charge from graphene \cite{kimFocusedLaserEnabledJunctionsGraphene2013}, it would form a dipolar layer with an electric field distribution deviating from our experiment results, as shown in the gray curve in Fig. \ref{fig:ElectricAnalysis} (c). Apart from the substrate, the only contact with the SLG flake was the diamond probe, forming a graphene/diamond heterojunction, which provides a pathway for the charge transfer. Probes used in the experiment had undergone terminal oxidation treatment, which leads to a complex surface chemical structure with multiple kinds of charge traps \cite{crawfordSurfaceTransferDoping2021}. The convincing data obtained from near-surface NV spin relaxation measurements and surface spectroscopy support a model attributing the surface characteristics to a system of disordered charge traps with electrons undergoing random hopping between them \cite{sangtawesinOriginsDiamondSurface2019,dwyerProbingSpinDynamics2022}. The dangling bonds of sp$^3$ at the step edges \cite{dwyerProbingSpinDynamics2022} and some primal sp$^2$ defects \cite{staceyEvidencePrimalSp22019} are possible candidates for these charge traps. With the knowledge about the oxidized diamond surface, we provide the following charge transfer mechanism.
	
During the scanning experiments, a thermalized hot electron gas is established in graphene under a continuous and focused laser \cite{tielrooijPhotoexcitationCascadeMultiple2013,silva-guillenElectronHeatingMechanical2020,pognaHotCarrierCoolingHighQuality2021}, some electrons in the hot tail of the thermalized distribution overcome the work function and enter electric traps on the diamond surface, this process is called photo-thermionic (PTI) emission \cite{massicottePhotothermionicEffectVertical2016}. The PTI emission promotes the electron density adjacent to the contact point, thus drives an outward diffusion of electrons on the diamond surface, as shown in Fig. \ref{fig:RealtimeTrace} (d). The diffusion further generates a built-in electric field, which in turn slows down the process till a new steady state forms in Fig. \ref{fig:RealtimeTrace} (e). Similar electron diffusion from SLG to diamond has been observed in a wide-field NV imaging experiment during which charge redistribution within several micrometers from SLG edges was reported \cite{lillieImagingGrapheneFieldEffect2019}. It is worth noting that due to the high electron occupancy rate in such a steady state, electron hopping should be suppressed by Coulomb blockade, this may weaken the surface electric screening that has been hampering the application of NV electric sensing \cite{obergSolutionElectricField2020}. Since the steady state depends on the balance between the laser-induced high electron density and the built-in electric field, once the continuous laser is turned off, a recombination driven by the built-in field will bring the system back to equilibrium in Fig. \ref{fig:RealtimeTrace} (f) and (g), which explains the exponential decay in Fig. \ref{fig:RealtimeTrace} (c). 
	
	
This work studied the electric edge effect of charged graphene flakes with nanoscale scanning NV microscopy. The FRET was utilized to determine positions of graphene edges and a quantitative analysis of the electric field distribution near edges was conducted and compared to a charged graphene model. The charge transfer mechanism was investigated and the recombination process was observed using real-time tracing methods. 

The charging mechanism and the edge effect of floating graphene flakes could be applied to research areas such as optical control of one-dimensional floating gates and photo-catalysis with graphene. In addition, the charge transfer process between the graphene and diamond may weaken the surface screening that limits NV electric quantum sensing with shallow NV, providing a new method for improving the NV electric sensitivity. Photo-thermionic diffusion observed here also provides a new way to control electron spin bath on diamond surface, this is helpful for understanding the chemical structure of diamond surfaces and improving near-surface spin coherence. 

\section*{Author contribution}
 J.D. and F.S. supervised the project; J.D., F.S. and Z.D. proposed the idea and designed the experiments. Z.S.C. performed the experiments; Z.D. conducted the simulations; X.L., J.J. and G.S. suggested on the theoritical explanation; X.F., W.Z., and C.Z. provided the graphene samples; J.F. and H.Z. fabricated the graphene with femotolaser; Y.W. prepared the diamond probe; Z.C. and Z.Y. prepared the electrode, Z.D., Z.S.C., Y.S., Y.L., K.Y. and P.W. built the setup; Z.D., Z.S.C., and F.S. wrote the manuscript. All authors discussed and analyzed the data.
	
\section*{Declarations}
	The authors declare no competing interests.

\begin{acknowledgments}
    
This work was supported by the National Natural Science Foundation of China (grant nos. T2125011, 92265204), the CAS (grant nos. GJJSTD20200001, YSBR-068, YSBR-049), Innovation Program for Quantum Science and Technology (Grant No. 2021ZD0302200, 2021ZD0303204), the China Postdoctoral Science Foundation (grant nos. 	GZC20232558), New Cornerstone Science Foundation through the XPLORER PRIZE, and the Fundamental Research Funds for the Central Universities.
	
This work was partially carried out at the USTC Center for Micro and Nanoscale Research and Fabrication.  
	
We thank Prof. Zhenhua Qiao for his valuable advices on the explanation of charge transfer mechanism. 
\end{acknowledgments}

\bibliographystyle{apsrev}
\bibliography{bibliography}

\end{document}